\documentclass[fleqn]{article}
\usepackage{espcrc2}
\usepackage{psfig}

\newcommand{\chicJ}{\chi_{cJ}}

\newcommand{\kk}{K^+K^-}

\newcommand{\pppp}{\pi^+\pi^-\pi^+\pi^-}
\newcommand{\ppkk}{\pi^+\pi^-K^+K^-}
\newcommand{\pppr}{\pi^+\pi^-p\bar{p}}
\newcommand{\kkkk}{K^+K^-K^+K^-}

\newcommand{\dphi}{\phi\phi}
\newcommand{\phikk}{\phi K^+K^-}

\newcommand{\gpppp}{\gamma \pi^+\pi^-\pi^+\pi^-}
\newcommand{\gppkk}{\gamma \pi^+\pi^-K^+K^-}

\newcommand{\gkkkk}{\gamma K^+K^-K^+K^-}

\newcommand{\psp}{\psi(2S)}
\newcommand{\jpsi}{J/\psi}
\newcommand{\ar}{\rightarrow}

\newcommand{\ww}{\omega\omega}

\newcommand{\bfg}{\begin{figure}}
\newcommand{\efg}{\end{figure}}
\newcommand{\bitm}{\begin{itemize}}
\newcommand{\eitm}{\end{itemize}}
\newcommand{\bnum}{\begin{enumerate}}
\newcommand{\enum}{\end{enumerate}}
\newcommand{\btbl}{\begin{table}}
\newcommand{\etbl}{\end{table}}
\newcommand{\btbu}{\begin{tabular}}
\newcommand{\etbu}{\end{tabular}}
\newcommand{\bcl}{\begin{center}}
\newcommand{\ecl}{\end{center}}
\newcommand{\bbt}{\bibitem}
\newcommand{\beq}{\begin{equation}}
\newcommand{\eeq}{\end{equation}}
\newcommand{\beqr}{\begin{eqnarray}}
\newcommand{\eeqr}{\end{eqnarray}}

\begin{document}
\pagestyle{empty}
\title{Measurements of $\chicJ \to \kkkk$ decays}
\author{\small{M.~Ablikim$^{1}$,     J.~Z.~Bai$^{1}$,       Y.~Ban$^{12}$,
J.~G.~Bian$^{1}$,              X.~Cai$^{1}$,               H.~F.~Chen$^{17}$,
H.~S.~Chen$^{1}$,              H.~X.~Chen$^{1}$,           J.~C.~Chen$^{1}$,
Jin~Chen$^{1}$
% ...
\vspace{0.2cm}\\
(BES Collaboration)\\
\vspace{0.2cm}
{\it
$^{1}$ Institute of High Energy Physics, Beijing 100049, People's Republic of China\\
% ...
$^{12}$ Peking University, Beijing 100871, People's Republic of China\\
% ...
$^{17}$ University of Hawaii, Honolulu, HI 96822, USA\\
% ...
}}}

\date{\today}

\begin{abstract}
Using 14M $\psi(2S)$ events taken with the BES-II detector, $\chicJ\ar
2(\kk)$ decays are studied. For the four-kaon final state, the
branching fractions are ${\cal B}(\chi_{c0,1,2}\ar 2(\kk))=(3.48\pm
0.23\pm 0.47)\times 10^{-3}$, $(0.70\pm 0.13\pm 0.10)\times 10^{-3}$,
and $(2.17\pm 0.20\pm 0.31)\times 10^{-3}$. For the $\phikk$ final
state, the branching fractions, which are measured for the first time, are
${\cal B}(\chi_{c0,1,2}\ar\phikk)=(1.03\pm 0.22\pm 0.15)\times 10^{-3},
(0.46\pm 0.16\pm 0.06)\times 10^{-3}$, and $(1.67\pm 0.26\pm 0.24)\times
10^{-4}$. For the $\dphi$ final state, ${\cal
B}(\chi_{c0,2}\ar\dphi)=(0.94\pm 0.21\pm 0.13)\times 10^{-3}$ and
$(1.70\pm 0.30\pm 0.25)\times 10^{-3}$.
\end{abstract}
\maketitle
\section{Introduction}
Exclusive quarkonium decays provide an important laboratory for investigating 
perturbative quantum chromodynamics. Compared with $\jpsi$ and $\psi(2S)$ 
decays, there is much less knowledge on $\chicJ$ decays which have
parity and charge conjugation $PC=++$. 
Relatively few exclusive decays of the $\chicJ$ have been 
measured.
For the $\chicJ\ar$ Vector Vector (VV) mode, measurements of
$\chicJ\ar\phi\phi$~\cite{2phi}, $K^*(892)^0\bar{K}^*(892)^0$~\cite{kstar}, 
and $\ww$~\cite{ww} have been recently reported.  The search for new 
decay modes and measurements with higher precision will 
help in better understanding various $\chicJ$ decay
mechanisms~\cite{color,hrb} and the nature of $^3P_J~c\bar{c}$ bound
states.

 Furthermore, the decays of $\chicJ$, especially $\chi_{c0}$ and
$\chi_{c2}$, provide a direct window on glueball dynamics in the
$0^{++}$ and $2^{++}$ channels since the hadronic decays may proceed
via $c\bar{c}\ar gg\ar q\bar{q}q\bar{q}$. Recently, a paper by
Zhao~\cite{zhaoq} points out that the decay branching fractions for scalar 
glueball candidates ($f_0(1370)$, $f_0(1500)$, and $f_0(1710)$) in $\chi_{c0}$ 
decays may be predicted by a factorization scheme, in which some
parameters can be fitted with ${\cal B}(\chicJ\ar\ww,
K^*(892)^0\bar{K}^*(892)^0,\dphi)$~\cite{zhaoq}.  The measurement
precision of ${\cal B}(\chicJ\ar VV)$ will affect the uncertainties of
the fitted parameters. Also, these fitted parameters
will help clarify the role played by OZI-rule violation
and SU(3) flavor breaking in the decays. 

In this analysis, $\chicJ \to \kkkk$ is studied
 using $\psp$ radiative decays. The branching 
fractions of $\chicJ\ar\kkkk$ and $\chi_{c0,2}\ar\dphi$ are measured 
with higher statistics, and those of $\chicJ$ decaying 
to $\phikk$ are measured for the first time.

\section{The BES detector}
The Beijing Spectrometer (BES) is a conventional solenoidal magnet
detector that is described in detail in Ref.~\cite{bes}; BESII is the
upgraded version of the BES detector~\cite{bes2}. A 12-layer vertex
chamber (VC) surrounding the beam pipe provides trigger
and position information. A forty-layer main drift chamber (MDC),
located radially outside the VC, provides trajectory and energy loss
($dE/dx$) information for charged tracks over $85\%$ of the total
solid angle.  The momenta resolution is $\sigma _p/p = 0.017
\sqrt{1+p^2}$ ($p$ in $\hbox{\rm GeV/c}$), and the $dE/dx$ resolution
for hadron tracks is $\sim 8\%$.  An array of 48 scintillation
counters surrounding the MDC measures the time-of-flight (TOF) of
charged tracks with a resolution of $\sim 200$ ps for hadrons.
Outside of the TOF counters is a 12-radiation-length barrel shower
counter (BSC) composed of gas tubes interleaved with
lead sheets. This measures the energies of electrons and photons over
$\sim 80\%$ of the total solid angle with an energy resolution of
$\sigma_E/E=22\%/\sqrt{E}$ ($E$ in GeV).  Outside of the solenoidal
coil, which provides a 0.4~Tesla magnetic field over the tracking
volume, is an iron flux return that is instrumented with three double
layers of counters that identify muons of momenta greater than 0.5
GeV/c.

A GEANT3 based Monte Carlo (MC) program with
detailed consideration of the detector performance (such as dead
electronic channels) is used to simulate the BESII detector.  The
consistency between data and Monte Carlo has been carefully checked in
many high purity physics channels, and the agreement is quite
reasonable~\cite{simbes}.

\section{Event selection}
The data sample used for this analysis consists of $(14.00\pm 
0.56)\times 10^6~\psp$ events ~\cite{moxh} collected with the BESII detector 
at the center-of-mass energy $\sqrt s=M_{\psp}$. The $\chicJ\ar\kkkk$ 
channels are investigated using $\psi(2S)$ radiative decays 
to $\chicJ$. Events with four charged tracks and one to three photons are 
selected. Each charged track is required to be well fitted by a helix and to 
have a polar angle, $\theta$, within the fiducial region $|\cos\theta|<0.8$. 
To ensure tracks originate from the interaction region, we require 
$V_{xy}=\sqrt{V_x^2+V_y^2}<2$ cm and $|V_z|<20$ cm, where $V_x$, $V_y$, 
and $V_z$ are the $x, y$, and $z$ coordinates of the point of closest 
approach of each charged track to the beam axis. All charged tracks
must be identified as kaons using the combined dE/dx and 
TOF information.  

A neutral cluster is considered to be a photon candidate if it is
located within the BSC fiducial region ($|\cos\theta|<0.75$), the
energy deposited in the BSC is greater than 50 MeV, the first hit
appears in the first 6 radiation lengths, the angle between the
cluster and the nearest charged track is more than $15^\circ$, and
the angle between the direction of cluster development and the
direction of the photon emission is less than $40^\circ$.

A four constraint (4-C) kinematic fit under the $\psp\ar\gkkkk$ hypothesis
is performed, and the $\chi^2$ of the fit is required to be less
than 25. For events with two or three photon 
candidates, the combination having the minimum $\chi^2$ is chosen. 
In addition, $\chi^2_{\gamma\kkkk}<\chi^2_{\gamma\pppp}$ and
$\chi^2_{\gamma\kkkk}<\chi^2_{\gamma\ppkk}$ are required to
suppress background contamination from $\psp\ar\gpppp$ and
$\psp\ar\gppkk$. Figure~\ref{chi} shows the $\chi^2$ distribution of 
data and MC for the process of $\chi_{c0}\ar\kkkk$.
\bfg
\centerline{\psfig{file=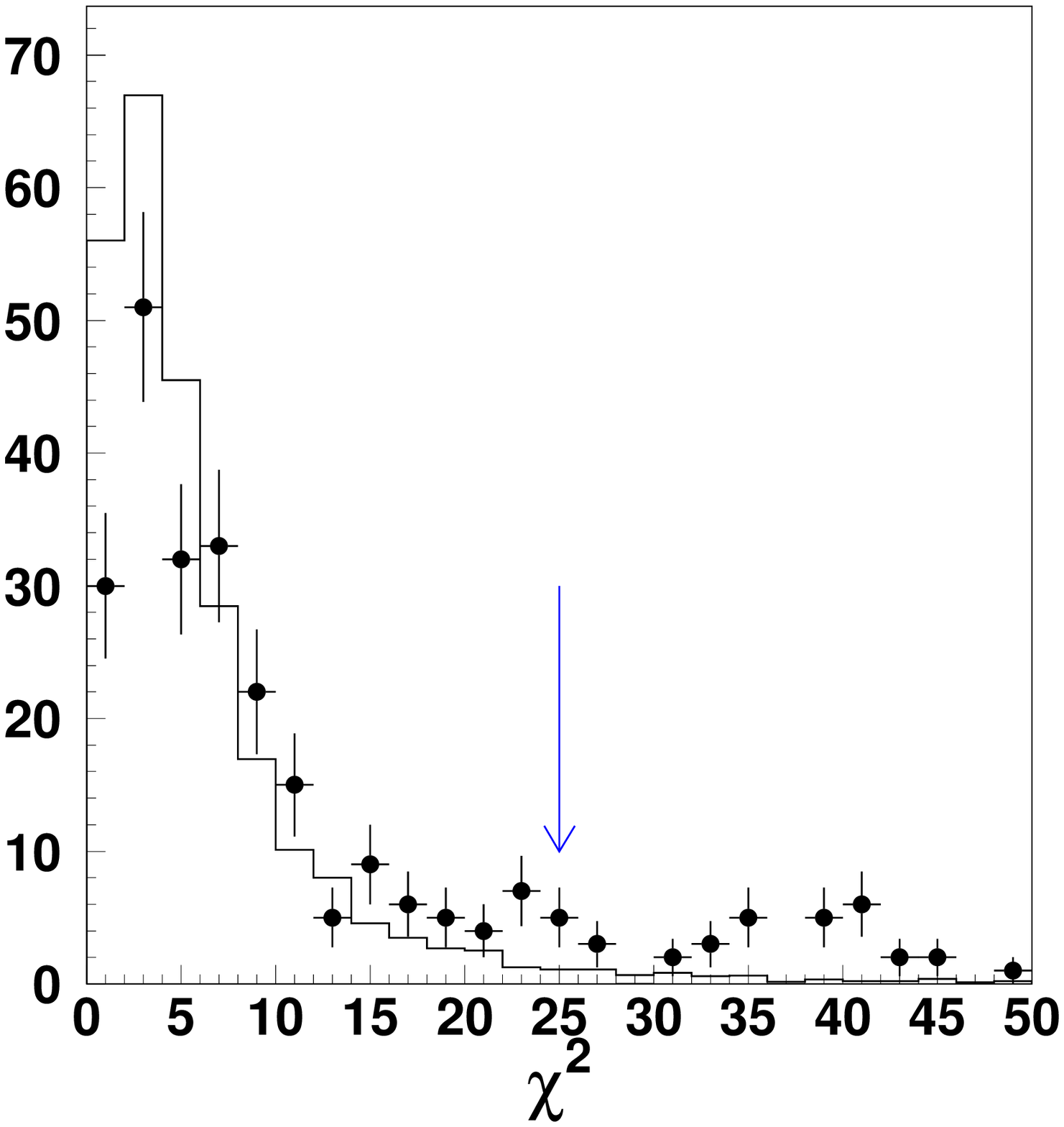,width=6cm,height=5.5cm}}
\caption{$\chi^2$ distribution of 4-C fit for $\chi_{c0}\ar\kkkk$. 
Histogram denotes MC and error bars denote data.}
\label{chi}
\efg

With four selected kaons, there are four ways to combine oppositely
charged kaons, and two combinations of $M_{\kk}^{(1)}~M_{\kk}^{(2)}$
pairs can be formed. The combination that has one of its
$M_{\kk}$ closest to the $\phi$ mass is selected for further analysis.

Figure~\ref{scatter} shows the distribution of $M_{\kk}$ versus
$M_{\kk}$ for selected events. There are two clear bands near 1.02
GeV/$c^2$ which correspond to the $\phikk$ final state. The insert in
the upper right corner, is the enlarged view of the lower left corner,
and a clear $\dphi$ signal can be seen.  Figure~\ref{mcscatter} shows
the corresponding $M_{\kk}^{(1)}$ versus $M_{\kk}^{(2)}$ distribution for MC
$\psp\ar\gamma\chi_{c1}$, $\chi_{c1}\ar\phikk$ and $\dphi$, using ${\cal
B}(\chi_{c1}\ar\phikk):{\cal B}(\chi_{c1}\ar\dphi)$=2:1.  
\bfg
\centerline{\psfig{file=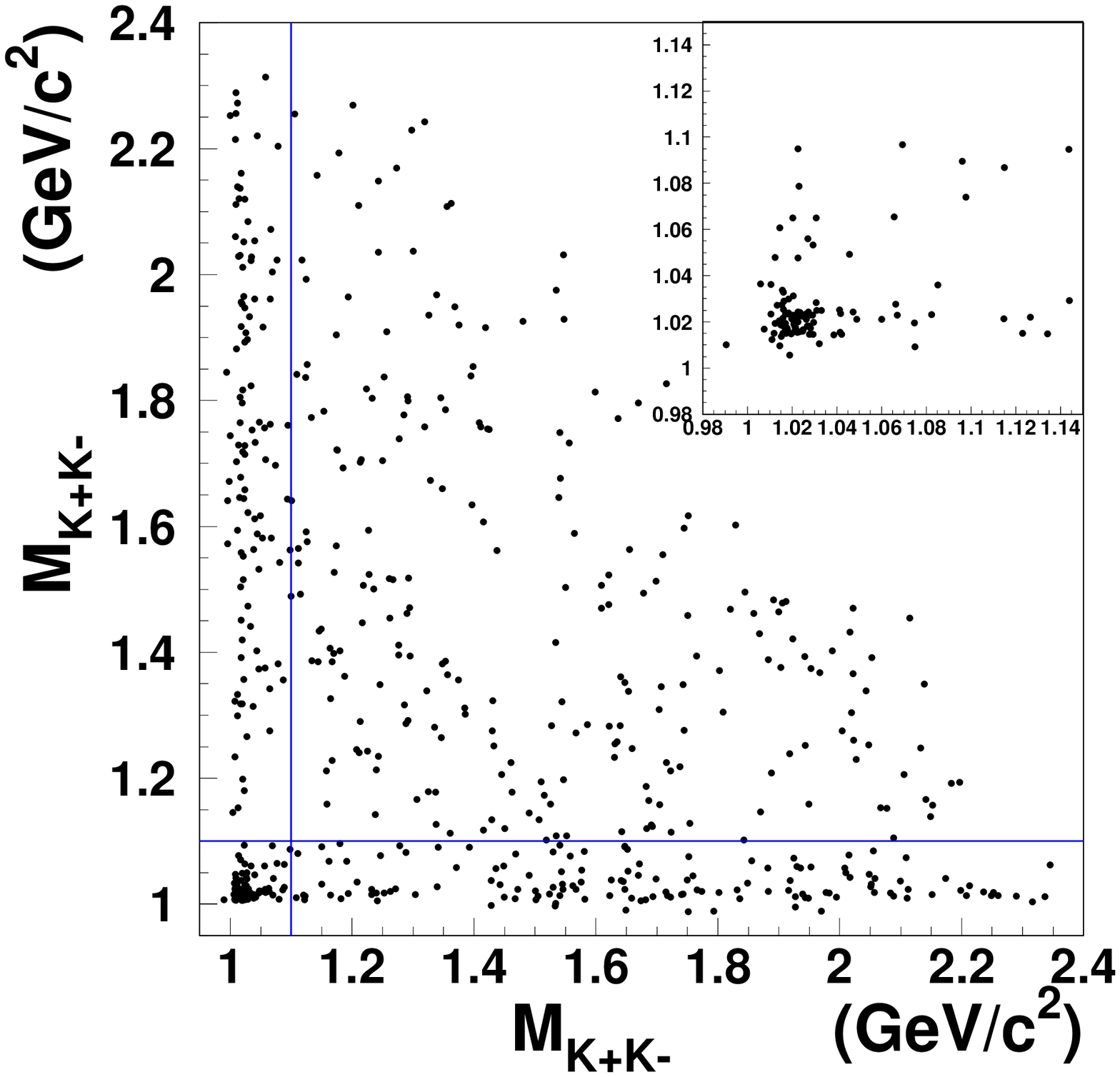,width=6cm,height=5.5cm}}
\caption{$M_{\kk}^{(1)}$ versus $M_{\kk}^{(2)}$ for candidate events.}
\label{scatter}
\efg

\bfg
\centerline{\psfig{file=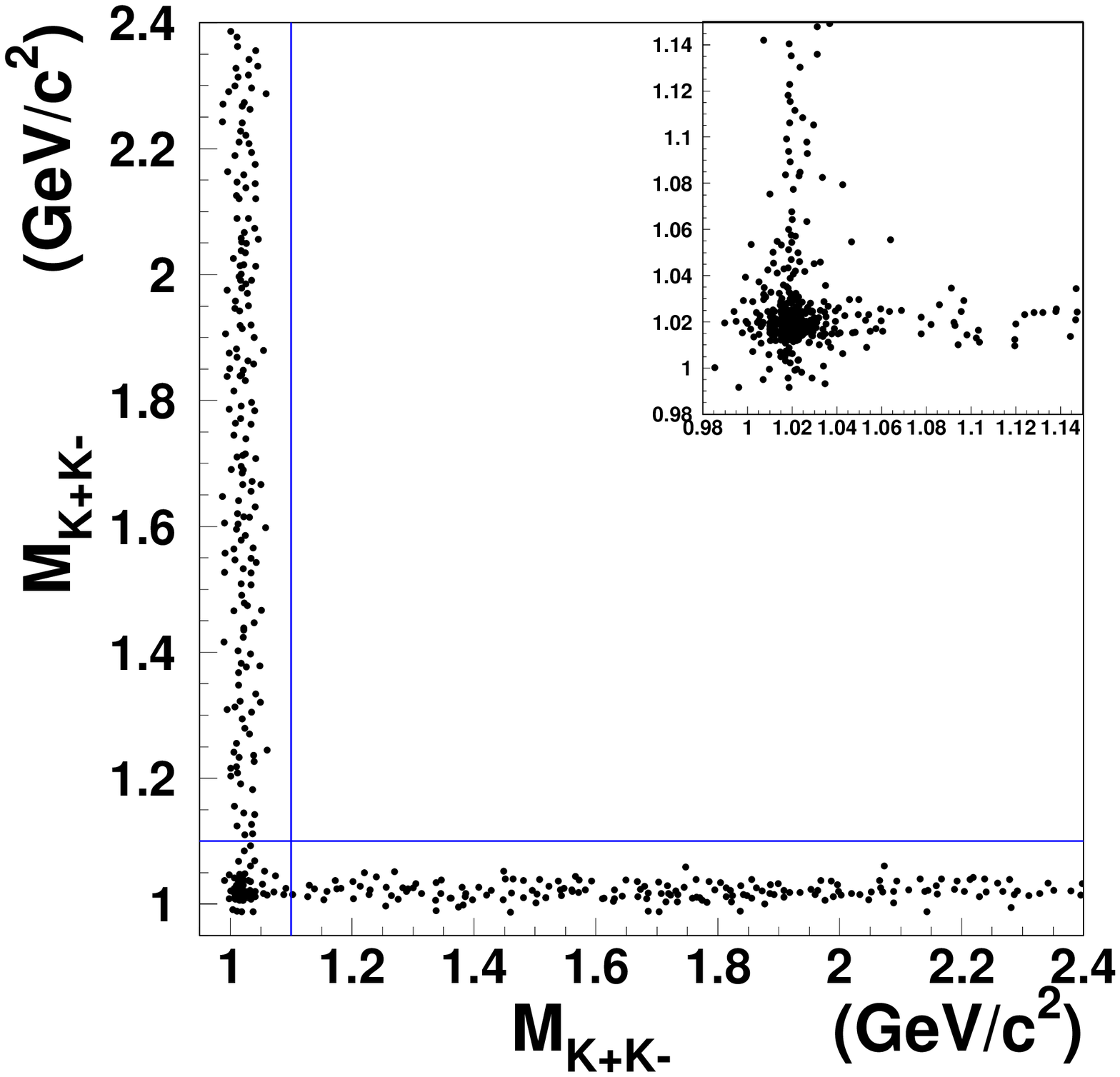,width=6cm,height=5.5cm}}
\caption{$M_{\kk}^{(1)}$ versus $M_{\kk}^{(2)}$ for MC 
$\psp\ar\gamma\chi_{c1}$,
  $\chi_{c1}\ar\phikk$ and $\dphi$ events after event selection.}
\label{mcscatter}
\efg

 To investigate intermediate resonances in $(\kk)$ final states, the
 invariant mass distribution of all four possible $\kk$ combinations
 are plotted in Figure~\ref{allkk}(a). Except for the $\phi$, no other
 obvious resonance is seen.
\bfg[hb]
\centerline{\psfig{file=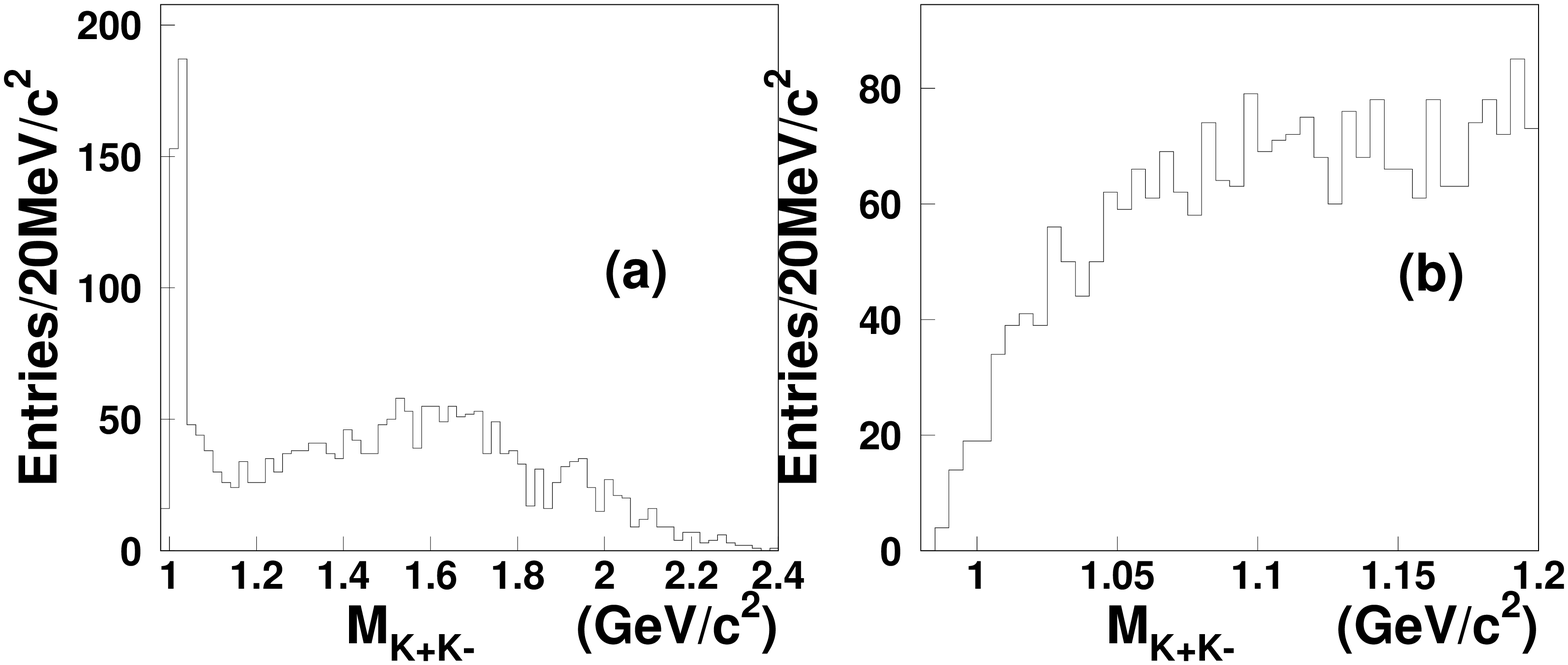,width=8cm,height=4cm}}
\caption{$M_{\kk}$ distribution of all four possible $\kk$ combinations.
(b) $M_{\kk}$ distribution for MC $\psp\ar\gamma\chi_{c1},\chi_{c1}\ar\kkkk$.}
\label{allkk}
\efg

To test if the above selection criteria will cause ``fake'' $\phi$
signals, $\psp\ar\gamma\chi_{c1}, \chi_{c1}\ar\kkkk$ MC events are
generated according to phase space. Figure~\ref{allkk}(b) shows the
$M_{\kk}$ distribution of these events using the same selection as for
data. No peak is seen around the $\phi$ signal region.

\section{MC simulation}
For each of the channels studied 100,000 MC events are generated. The
proper angular distributions for the photons emitted in
$\psi(2S)\ar\gamma\chicJ$ are used~\cite{trans}. Phase space is used
for the $\chicJ\ar 2(\kk)$ decays (including intermediate states,
e.g. $\chicJ\ar\dphi$).

\section{Background study}
Possible backgrounds come from $\psp\ar\gamma\chicJ$,
~$\chicJ\ar\pppp$,~$\ppkk$,~$K_sK\pi$, and $\pppr$;
$\psp\ar\pi^0\kkkk$; and phase space $\psp\ar\gkkkk$. 100,000 MC
events are generated for each of the first four background channels
where the angular distribution for the radiative photon is generated
the same as for the signal channels while the $\chicJ$ decays are
generated according to phase space. The contamination from these
backgrounds to each signal channel is less than 1.0\%, which is
negligible. 100,000 MC events are also generated for each of the last
two background channels according to phase space. For
$\psp\ar\pi^0\kkkk$, Figure~\ref{bkgshape}(a) shows the $M_{\kkkk}$
distribution in the region 3.2-3.7 GeV/$c^2$. Using the branching
fraction measured by CLEOc~\cite{13chan}, the number of events from
this channel is expected to be $23\pm 7$.  For $\psp\ar\gkkkk$, the
$M_{\kkkk}$ distribution is shown in Figure~\ref{bkgshape}(b). This
branching fraction is currently unavailable. However, comparing the
$M_{\kkkk}$ distribution from this channel and that from data, shown
in Figure~\ref{4kfit}, the contribution from $\psp\ar\gkkkk$ background 
should be small. No $\chicJ$ peaks are seen in Figure~\ref{bkgshape}(a) and 
(b).  Thus, the fitted number of $\chicJ$ signal events 
is insensitive to the shape of the function used to describe the total 
background.

We have also considered possible background from
$\psp\ar\gamma\chi_{c0},\chi_{c0}\ar f_0(980)f_0(980)\ar\kkkk$. Using
the branching ratio for $\chi_{c0}\ar f_0(980)f_0(980)\ar \pi^+ \pi^-
K^+ K^-$
\cite{pwachi0}, the ratio $B(f_0(980)\ar K^+ K^-)/(B(f_0(980)\ar K^+ K^-)
+ B(f_0(980)\ar \pi^+ \pi^-))$~\cite{pwachi0}, and MC efficiencies for
these processes, the estimated number of $f_0(980)f_0(980)\ar\kkkk$
events is about 6, which are spread over a relatively wide region
compared with the width of the $\phi$ ~\cite{phikk}.  Thus, this
background in the $\phi$ region is negligibly small.
\bfg
\centerline{\psfig{file=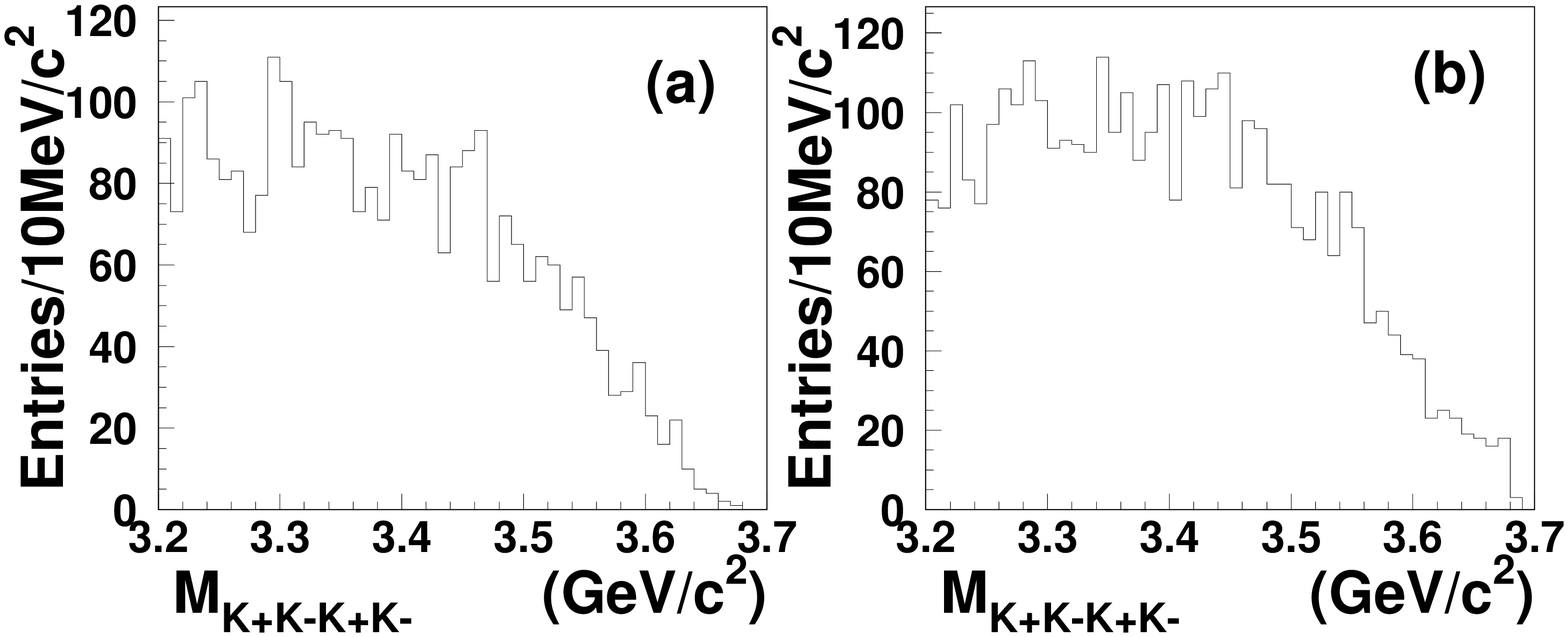,width=8cm,height=4.5cm}}
\caption{Background shape obtained from MC simulation.
(a) $\psp\ar\pi^0\kkkk$. (b) $\psp\ar\gkkkk$.} 
\label{bkgshape}
\efg
\section{Mass spectrum fit}
\subsection{\boldmath $\chicJ\ar 2(\kk)$}
Figure~\ref{4kfit}(a) shows the  high mass ($>3.2$ GeV/$c^2$) $2(\kk)$
invariant mass distribution using all candidate events in 
$\psp\ar\gamma\chicJ,\chicJ\ar 2(\kk)$.
Clear $\chicJ$ signals can be seen in this Figure. A fit with
Breit-Wigner functions convoluted with 
Gaussian resolution functions (about 15 MeV/$c^2$ for $\chicJ$ signals in all 
channels studied) yields the number of  $\chicJ$ events:
$N_{\chi_{c0}}=278\pm 18$, $N_{\chi_{c1}}=54\pm 10$, and 
$N_{\chi_{c2}}=160\pm 14$.
\bfg
\centerline{\psfig{file=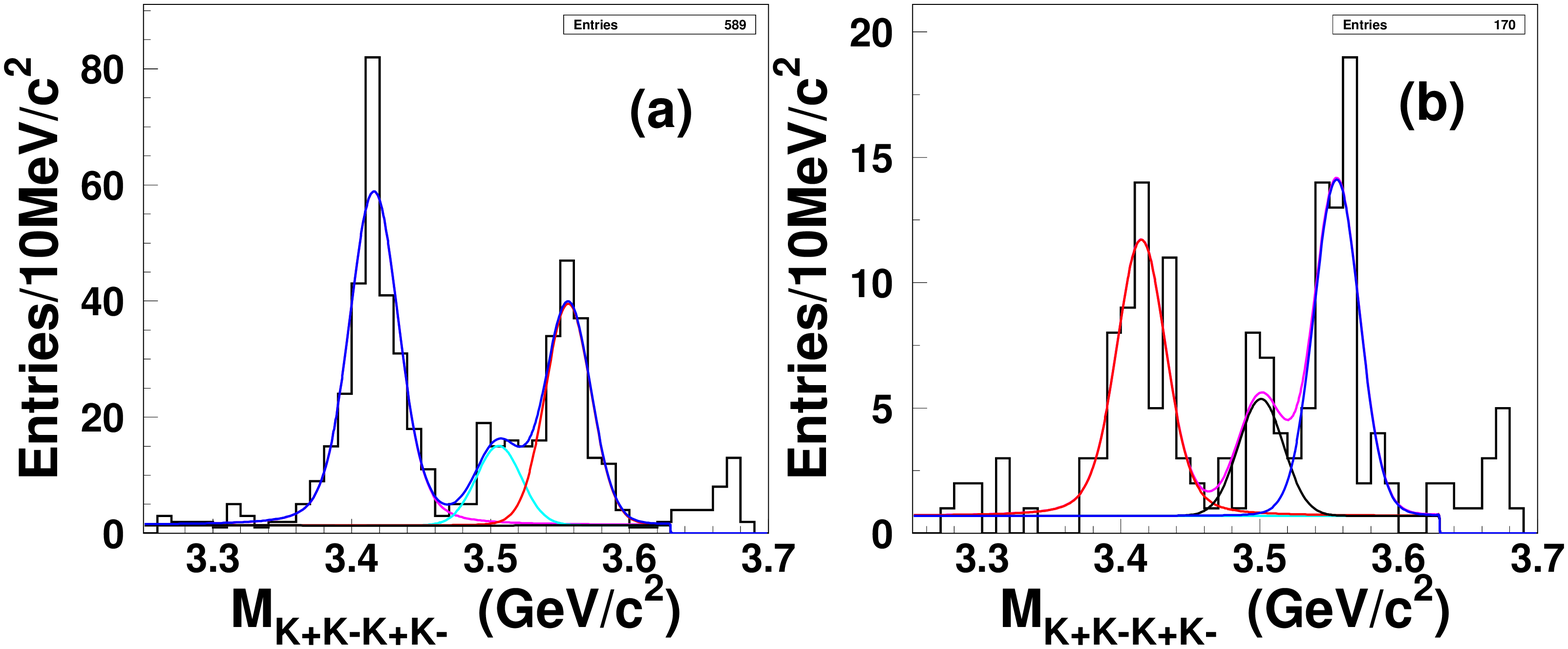,width=8cm,height=4.5cm}}
\caption{Breit-Wigner fit to $\chicJ$ signals (a) with all candidate
  events and (b) with the $\phikk$ region events.}
\label{4kfit}
\efg
MC simulation gives detection efficiencies of 5.9\%, 6.3\%, and 5.8\% 
for $J=0$, 1, and 2, respectively. 
\subsection{\boldmath $\chicJ\ar \phikk$}
The  $\chicJ\ar\phikk$ events are clustered as two bands around the $\phi$ 
mass in Figure~\ref{scatter}. The  $\phikk$ region is defined by
$M_{\kk}^{(i)}\in(1.08,2.4)$ GeV/$c^2$ and 
$M_{\kk}^{(j\ne i)}\in (1.00,1.04)$ GeV/$c^2$ ($i$, $j$ = 1, 2).
Figure~\ref{4kfit}(b) shows the $M_{\phikk}$ distribution for these events, 
and clear $\chicJ$ signals can be seen. MC simulation gives detection 
efficiencies of 5.9\%, 6.2\%, and 5.6\% for J=0, 1, and 2,
respectively. A fit with Breit-Wigner functions 
convoluted with Gaussian resolution functions yields the number of $\chicJ$ events:
$N_{\chi_{c0}}=53.5\pm 8.3$, $N_{\chi_{c1}}=19.2\pm 5.6$, and
$N_{\chi_{c2}}=56.3\pm 8.2.$

In order to determine the contribution from non-resonant
$\psp\ar\gamma\chicJ,\chicJ\ar 2(\kk)$ events in the
$\phikk$ region, we analyze the non-resonant $2(\kk)$
region, defined by $M_{\kk}^{(i)}\in(1.1,2.4)$ GeV/$c^2$ ($i$ = 1,2)
in Figure~\ref{scatter}. The number of events
in this region is $N_t^{nr}=238$.  Figure~\ref{msb}(a) shows the
$M_{\kkkk}$ distribution for the non-resonant $2(\kk)$ region events, where
three clean $\chi_{cJ}$ peaks are seen with very little background. 
A fit with Breit-Wigner functions convoluted with Gaussian resolution
functions yields: $N_{\chi_{c0}^{nr}}=144.1\pm 12.7$,
$N_{\chi_{c1}^{nr}}=21.5\pm5.5$, and $N_{\chi_{c2}^{nr}}=39.4\pm 6.9.$

Figure~\ref{msb}(b) shows the
$\phi$ signal for the $\phikk$ region events. The fit yields $143\pm 14$
signal events and $N_{bg}=25$ background events from non-resonant
$2(\kk)$ events within a 3$\sigma$
window (1.00 - 1.04 GeV/$c^2$) in the $\phi$ signal region. 
We calculate the  non-resonant
$2(\kk)$ contribution to $N_{\chi_{c0}}$, $N_{\chi_{c1}}$, and
$N_{\chi_{c2}}$ assuming that the proportion of $\chi_{c0}$,
$\chi_{c1}$, and $\chi_{c2}$ events is the same for non-resonant
$2(\kk)$ events in the non-resonant and the  $\phikk$ regions.
For $\chi_{c0}$, the non-resonant contribution is
$N_{\chi_{c0}}^{n}=N_{\chi_{c0}}^{nr}\cdot N_{bg}/N_t^{nr}=15.1\pm
1.3$. Similarly, we obtain $N_{\chi_{c1}}^{n}=2.3\pm0.6$, and
$N_{\chi_{c2}}^{n}=4.1\pm 0.7$.  Therefore, the number
of  $\phikk$ events, after subtracting the non-resonant
$2(\kk)$ events, is $N_{\chi_{c0}}=38.4\pm 8.4$,
$N_{\chi_{c1}}=16.9\pm 5.6$, and $N_{\chi_{c2}}=52.2\pm 8.2.$
\bfg
\centerline{\psfig{file=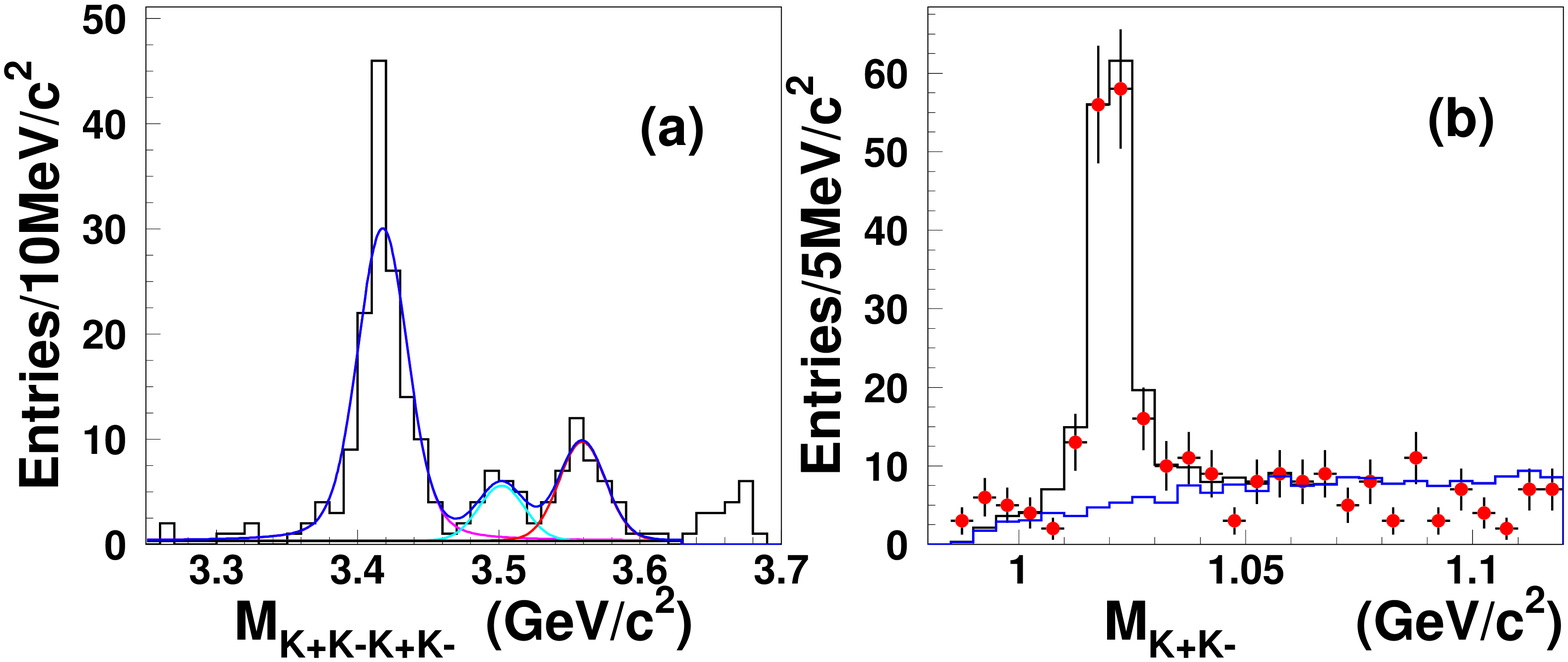,width=8cm,height=4.5cm}}
\caption{(a) Breit-Wigner fit to
$\chicJ$ using non-resonant $2(\kk)$ region events. (b) Fit to $\phi$
signal with $\phikk$ region events.}
\label{msb}
\efg

\subsection{\boldmath $\chicJ\ar\dphi$}
The signal region for $\chicJ\ar\dphi$ events is a  40 MeV/$c^2\times$
40 MeV/$c^2$ square 
around the $\phi$ mass in Figure~\ref{scatter}. 
Figure~\ref{m2phi} shows the $M_{\dphi}$ distribution for these $\dphi$ 
events, and clear $\chi_{c0,2}$ signals can be seen. MC simulation gives 
detection efficiencies of 9.0\%  and 8.6\% for J=0 and 2,
respectively. A fit yields
the number of events for $\chicJ$:
$N_{\chi_{c0}}=27.8\pm 5.8$ and $N_{\chi_{c2}}=42.7\pm 7.1.$
\bfg
\centerline{\psfig{file=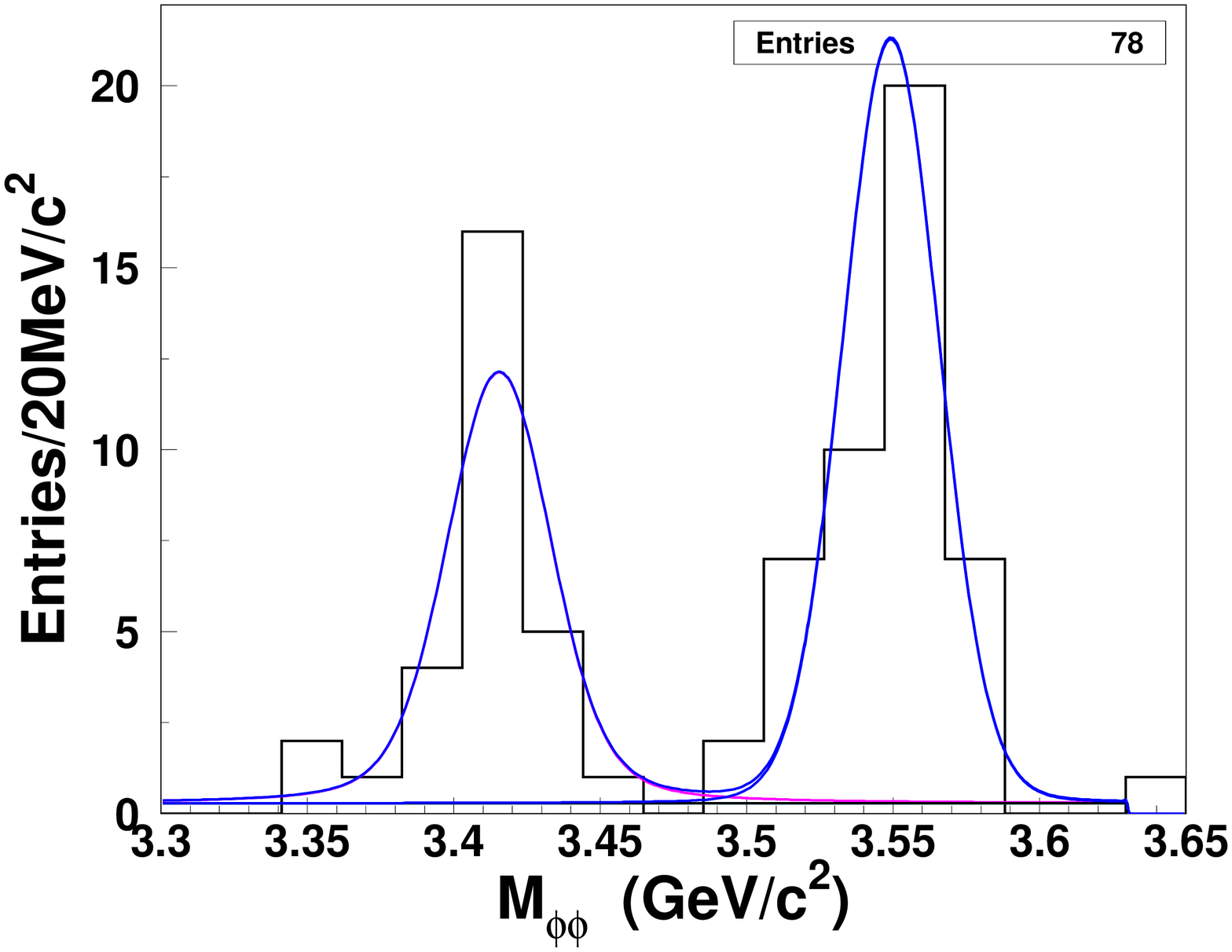,width=6cm,height=5cm}}
\caption{Fit to $\chicJ$  signals in $\dphi$ final state.}
\label{m2phi}
\efg

The two bands near the $\phi$ mass in Figure~\ref{scatter}, used to 
extract the $\chicJ\ar\phikk$ signal in Section 6.2, are taken as the
sideband region for 
the $\dphi$ events. They include both the $\phikk$ events and
non-resonant $K^+ K^-K^+ K^-$
events.  From MC simulation the event distributions in the two bands
are nearly uniform. The number of normalized sideband events in the 
$\dphi$ signal region are
$$N_{\chi_{c0}}^{sd}=\frac{53.5\pm 8.3}{f}\approx1.6\pm 0.3,$$
~$$N_{\chi_{c2}}^{sd}=\frac{56.3\pm 8.2}{f}\approx1.7\pm 0.3,$$
respectively, where the factor f=33 is the ratio of the sideband area
to the $\dphi$ signal region area.  Thus we obtain the 
number of events in $\chicJ\ar\dphi$:
$N_{\chi_{c0}}=26.2\pm 5.8$ and  $N_{\chi_{c2}}=41.0\pm 7.1.$
\section{Systematic error}
The systematic error in these branching fraction measurements includes the
uncertainties caused by wire resolution, particle ID, photon
efficiency, and the 
number of $\psi(2S)$ events.

The systematic error caused by MDC tracking and the kinematic fit are
estimated by using simulations with different MDC wire resolutions
~\cite{simbes}.  For particle ID, the combined information of $dE/dx$
and TOF is used. An error of 2\% is assigned for each charged
track~\cite{wangwf} and each photon~\cite{simbes}. The errors introduced
by branching fractions of intermediate states are taken from the Particle Data 
Group (PDG)~\cite{pdg}. 

The total systematic errors, determined by the sum of all sources
added in quadrature, are listed in Table~\ref{tot}. The uncertainty from
${\cal B}(\phi\ar\kk)$ contributes once  in the
systematic error estimation for $\chicJ\ar\phikk$ and twice in $\dphi$,
while it does not
contribute in $\chicJ\ar2(\kk)$.
For the uncertainties caused by wire resolution, there are
some slight differences for the different decay channels.
\begin{table*}
\caption{Systematic error (\%). In the  wire resolution row, the
numbers from left to right correspond to $\psp\ar 2(\kk)$, $\phikk$, and $\dphi$.}
\bcl
\doublerulesep 2pt
\begin{tabular}{l|c|c|c|c}
\hline\hline
\multicolumn{2}{l}{Source}\vline&$\chi_{c0}$&$\chi_{c1}$&$\chi_{c2}$\\\hline\hline
\multicolumn{2}{l}{Wire resolution}\vline&8.9, 9.8, 10.0&9.3, 9.9
&9.7, 9.6, 10.1\\
\multicolumn{2}{l}{Particle ID}\vline&8&8&8\\
\multicolumn{2}{l}{Photon efficiency}\vline&2&2&2\\
\multicolumn{2}{l}{Background shape}\vline&negligible&negligible&negligible\\
\multicolumn{2}{l}{Number of $\psi(2S)$}\vline&4&4&4\\
\multicolumn{2}{l}{${\cal B}(\psi(2S)\ar\gamma\chicJ)$}\vline&4.3&4.6&4.9\\
\multicolumn{2}{l}{${\cal B}(\phi\ar\kk)$}\vline&1.2&1.2&1.2\\\hline
&$\chicJ\ar2(\kk)$&13.5&13.9&14.3\\
Total&$\chicJ\ar\phikk$&14.1&14.3&14.2\\
&$\chicJ\ar\dphi$&14.3&-&14.5\\\hline
\end{tabular}
\label{tot}
\ecl
\end{table*}
\section{Results}
For $\chicJ\ar 2(\kk)$ (including intermediate states), the branching 
fractions are calculated using
$$
{\cal B}(\chicJ\ar2(\kk))=\\
\frac{N_{\chicJ}}{N_{\psi(2S)}\cdot{\cal
B}(\psi(2S)\ar\gamma\chicJ)\cdot\bar{\epsilon}},
$$
where, the average detection efficiency $\bar{\epsilon}$ is given by
$$\bar{\epsilon}=\frac{N_{\chicJ}-N_{\phikk}-
N_{\dphi}}{N_{\chicJ}}\cdot\epsilon_{2(\kk)}+$$
$$\frac{N_{\phikk}}{N_{\chicJ}}\cdot\epsilon_{\phikk}+
\frac{N_{\dphi}}{N_{\chicJ}}\cdot\epsilon_{\dphi}.
$$
Similarly, we can calculate the branching fractions for
$\chicJ\ar\phikk$, $\dphi$ 
with corresponding efficiency expressions. Table~\ref{results} lists our 
measurement results, together with the PDG values.
\begin{table*}[hbtp]
\doublerulesep 0.5pt
\caption{\label{results} $\chicJ\ar 2(\kk)$ branching fractions.}
\vskip 0.2 cm
\begin{center}
\begin{tabular}{c|c|c|c|c|c}              \hline\hline
Channel&\multicolumn{2}{c}{$2(\kk) (\times 10^{-3})$}\vline
&$\phikk (\times 10^{-3})$&\multicolumn{2}{c}{$\dphi (\times 10^{-3})$}\\
\cline{2-3}\cline{4-5}\cline{5-6}
 &BES-II&PDG&BES-II&BES-II&PDG\\\hline
$\chi_{c0}$&$3.48\pm 0.23\pm 0.47$&$2.1\pm
0.4$&$1.03\pm 0.22\pm 0.15$&$0.94\pm 0.21\pm0.13$&$0.9\pm 0.5$\\\hline
$\chi_{c1}$&$0.70\pm 0.13\pm 0.10$&$0.39\pm
0.17$&$0.46\pm 0.16\pm 0.06$&$-$&$-$\\\hline
$\chi_{c2}$&$2.17\pm 0.20\pm 0.31$&$1.41\pm
0.35$&$1.67\pm 0.26\pm 0.24$&$1.70\pm 0.30\pm 0.25$&$1.9\pm 0.7$\\\hline
\end{tabular}
\end{center}
\end{table*}

In summary, the decays of $\chicJ\ar 2(\kk)$ are studied, and the
corresponding branching fractions including intermediate states are
given. The decay $\chicJ\ar\phikk$ is observed for the first time. The
branching fractions for $\chicJ\ar 2(\kk)$ and $\chicJ\ar\dphi$ are
measured with higher precision; Table~\ref{results} lists the
comparison of the measured branching fractions between BESII and the PDG.
Our measurement for $\chicJ\ar\dphi$, together with the two measurements of
$\chicJ\ar \ww$ and $\bar{K}^*(892)^0K^*(892)^0$, will be helpful in
understanding the nature of $\chicJ$ states.
\section{Acknowledgments}
The BES collaboration thanks the staff of BEPC and computing center for their 
hard efforts. This work is supported in part by the National Natural Science 
Foundation of China under contracts Nos. 10491300, 10225524, 10225525, 
10425523, the Chinese Academy of Sciences under contract No. KJ 95T-03, the 
100 Talents Program of CAS under Contract Nos. U-11, U-24, U-25, and the 
Knowledge Innovation Project of CAS under Contract Nos. U-602, U-34 (IHEP), 
the National Natural Science Foundation of China under Contract No.
10225522 (Tsinghua University), the Swedish research Council (VR), and the 
Department of Energy under Contract No.DE-FG02-04ER41291 (U Hawaii).
      
\end{document}